\documentclass[twocolumn,aps,superscriptaddress,multicol,amsmath,amssymb,showpacs]{revtex4-2}
\usepackage[utf8]{inputenc}
\usepackage{amsmath}
\usepackage{times}
\usepackage{siunitx}
\usepackage{graphicx}
\usepackage{sidecap}
\usepackage[dvipsnames]{xcolor}
\graphicspath{{.}{./figures-revs/}}
\usepackage{nameref,zref-xr}
\zxrsetup{toltxlabel}
\usepackage[colorlinks,linkcolor=blue,citecolor=blue,anchorcolor=blue]{hyperref}
\usepackage{cleveref}
\usepackage{textcomp}
\definecolor{dark-green}{rgb}{0.000, 0.500, 0.000}
\begin{document}

% \title{Competition between Jahn-Teller and breathing distortions in a (111)-oriented half-metallic manganite superlattice}
\title{
	Emergent half-metal with mixed structural order in (111)-oriented (LaMnO$_3$)$_{2n}\vert$(SrMnO$_3$)$_{n}$ superlattices
}

\author{Fabrizio Cossu}
\email[F. Cossu: ]{cossu@kangwon.ac.kr}
\affiliation{Department of Physics and Institute of Quantum Convergence \& Technology, Kangwon National University -- Chuncheon, 24341, Korea}
\affiliation{School of Physics, Engineering \& Technology, University of York, Heslington, York, YO10 5DD, United Kingdom}
%\affiliation{Asia Pacific Center for Theoretical Physics -- Pohang, 37673, Korea}

\author{J\`ulio Alves Do Nascimento}
%\email[J\`ulio Alves: ]{jcdn500@york.ac.uk}
\affiliation{School of Physics, Engineering \& Technology, University of York, Heslington, York, YO10 5DD, United Kingdom}

\author{Stuart A.\ Cavill}
%\email[Stuart A.\ Cavill: ]{stuart.cavill@york.ac.uk}
\affiliation{School of Physics, Engineering \& Technology, University of York, Heslington, York, YO10 5DD, United Kingdom}

\author{Igor Di Marco}
\affiliation{Institute of Physics, Faculty of Physics, Astronomy \& Informatics,
Nicolaus Copernicus University, Grudziadzka 5, 87-100, Toru\'n, Poland}
\affiliation{Department of Physics and Astronomy, Uppsala University, Box 516, SE-75120, Uppsala, Sweden}

\author{Vlado K.\ Lazarov}
%\email[Vlado Lazarov: ]{vlado.lazarov@york.ac.uk}
\affiliation{School of Physics, Engineering \& Technology, University of York, Heslington, York, YO10 5DD, United Kingdom}

\author{Heung-Sik Kim}
\email[H.-S. Kim: ]{heungsikim@kangwon.ac.kr}
\affiliation{Department of Physics and Institute of Quantum Convergence \& Technology, Kangwon National University -- Chuncheon, 24341, Korea}

%\author{Igor I.\ Mazin}
%\email{imazin2@gmu.edu}
%\affiliation{Department of Physics and Astronomy, George Mason University, Fairfax, VA, 22030, USA}
%\affiliation{Quantum Science and Engineering Center, George Mason University, Fairfax, VA, 22030, USA}

\date{\today}

\begin{abstract}
	 Using first-principles techniques, we study the structural, magnetic and electronic properties of (111)-oriented (LaMnO$_3$)$_{2n}$$\vert$(SrMnO$_3$)$_{n}$
	 superlattices of varying thickness ($n=2,4,6$). We find that the properties of the thinnest superlattice ($n=2$) are similar to the celebrated half-metallic
	 ferromagnetic alloy La$_{2/3}$Sr$_{1/3}$MnO$_3$, with quenched Jahn-Teller distortions. At intermediate thickness ($n=4$), the $a^{-}a^{-}a^{-}$ tilting pattern
	 transitions to the $a^{-}a^{-}c^{+}$ tilting pattern, driven by the lattice degrees of freedom in the LaMnO$_3$ region. The emergence of the Jahn-Teller modes
	 and the spatial extent needed for their development play a key role in this structural transition. For the largest thickness considered ($n=6$), we unveil an
	 emergent separation of Jahn-Teller and volume-breathing orders in the ground-state structure with the $a^{-}a^{-}c^{+}$ tilting pattern, whereas it vanishes in
	 the antiferromagnetic configurations. The ground state of all superlattices is half-metallic ferromagnetic, not affected by the underlying series of structural
	 transitions. Overall, these results outline a thickness-induced crossover between the physical properties of bulk La$_{2/3}$Sr$_{1/3}$MnO$_3$ and bulk LaMnO$_3$.
\end{abstract}

%\pacs{75.70.Cn}% insert suggested PACS numbers in braces on next line

\maketitle %\maketitle must follow title, authors, abstract and \pacs

\section{Introduction}
 Oxide thin films and superlattices hold great promise for future technologies, due to their remarkable versatility \cite{hoglund-Nature2022,jia_y-PRB.93.104403,yang_c-NanoL2023.8} and
 high-precision synthesis through advanced techniques such as molecular beam epitaxy \cite{prakash-wileybook.ch26,ismailbeigi-NatRevMat2017,roth_j-APLMat2021,lazarov_vk-PRL.107.056101}
 and pulsed laser deposition \cite{lorenz_m-wileybook.eap810,eres-PRL.117.206102,yao-apA2019,koster-JSNM2020,cheshire_d-JAP2022,gilks_d-SciRep2016}. Among them, manganites have been
 under remarkable attention for potential applications in oxide electronics and spintronics thanks to the ferromagnetic (FM) phase, a high spin polarisation and the emergence of colossal
 magnetoresistance both in the bulk \cite{sun_z-SciRep2013,prasad_b-AM2015.19,tokura-RPP2006,tokura-JMMM1999.1,dagotto-NJP2005,dagotto-PhysRep2001,baldini_m-PRL.106.066402} and in
 superlattices \cite{nakao-PRB.92.245104,miao_t-PNAS2020,gayathri_v-SciRep2023}. A major goal for the research on manganite superlattices is to reach ferromagnetism and half-metallicity
 at high temperatures \cite{haghirigosnet-JPhysD2003}. While the type of transport measurement determines whether true half-metallicity is observed \cite{nadgorny_b-PRB.63.184433} and 
 defects, spin-orbit coupling and temperature-dependent spin dynamics \cite{katsnelson-RMP.80.315} have a non-negligible effect, it is accepted that the prediction of half-metallicity
 from temperature-free models represent a valuable insight \cite{li_xx-NatSciRev2016,tang_qk-JMCC2022}. Early theoretical and experimental studies focussed on (001)-oriented mixed-valent
 manganite superlattices \cite{adamo-PRB.79.045125,bhattacharya-PRL.100.257203,nanda-PRB.79.054428,smadici-PRL.99.196404,dong-PRB.78.201102,chen_h-JPCM2017}; (111)-oriented superlattices
 with LaMnO$_3$ \cite{bruno-APLMat2017,zang_j-SciRep2017,gibert-nmat2012,wei_hm-APL2016,wang_zz-PSSb2022} or SrMnO$_3$ \cite{song_rn-FIP2015,wang_zz-PSSb2022} were also grown but remain
 underexplored due to difficulties in sample synthesis \cite{mantz-SS2020}, especially concerning the SrMnO$_3$ side \cite{wang_zz-PSSb2022}. Nevertheless, (111)-oriented superlattices
 can host intriguing properties due to their symmetric character \cite{ruegg-PRB.88.115146,ruegg-PRB.85.245131}, a polar discontinuity at the interface
 \cite{chakhalian-nmat2012,song_k-nnano2018,ryu_s-APLMat2022}, and a subtle competition between spin, orbital, charge, and lattice degrees of freedom.

 Common compounds are LaMnO$_3$ and SrMnO$_3$, respectively an orthorhombic ($Pnma$ space group) Jahn-Teller (J-T) insulator with A-type
 antiferromagnetic (AFM) coupling, $a^{-}a^{-}c^{+}$ tilting system (in Glazer's notation \cite{glazer-AC:B1972}) and Mott correlation and
 a cubic ($Pm\bar{3}m$) band insulator with G-type AFM coupling and negligible octahedral tilts. Their solid mixture with 1/3 Sr and 2/3
 La is a rhombohedral ($R\bar{3}c$ space group \cite{saleem-RSCAdv2018}) half-metal with FM coupling, $a^{-}a^{-}a^{-}$ tilting system
 and colossal magnetoresistance \cite{tokura-JMMM1999.1,tokura-jpsj1994,imada-RMP.70.1039}. The $Pnma$ space group with the $a^{-}a^{-}c^{+}$
 tilting system and the presence of the J-T distortions are crucial for the stability of the A-type AFM order of bulk LaMnO$_3$
 \cite{pickett_we-PRB.53.1146,geck-NJP2004,pavarini-PRL.104.086402,schmitt-PRB.101.214304}. Likewise other perovskite compounds, in
 superlattices we expect a competition of different tilting systems, charge and orbital orders, and various magnetic states; strain and
 stoichiometry provide a route to tune the tilting system \cite{myself-EPL2017,myself-npjCM2022} or to induce a crossover from orbital order
 to charge order \cite{schmitt-PRB.101.214304,myself-npjCM2022,mazin-PRL.98.176406}; magnetic phase transitions, coexistence or separation
 may also occur \cite{tebano-PRL.100.137401,pruneda-PRL.99.226101,congiu-JMMM2016,myself-PRB.87.214420}. \textit{Ab-initio} studies may be
 an outstanding instrument to determine the interplay of various degrees of freedom and predict emergent properties in superlattices. Our
 pilot study on a (111)-oriented (LaMnO$_3$)$_{12}\vert$(SrMnO$_3$)$_{6}$ superlattice with the $a^{-}a^{-}a^{-}$ tilting system revealed
 the presence of a robust half-metallic phase with rhombohedral symmetry that can be stabilized with a small in-plane compressive strain
 \cite{myself-npjCM2022}. In this article, we present the results of \textit{ab-initio} calculations of (LaMnO$_3$)$_{2n}\vert$(SrMnO$_3$)$_{n}$
 superlattices with $n=2,4,6$. We provide a complete overview of structural, electronic and magnetic properties against varying thickness,
 in the ground-state as well as in excited states \footnote{Notice that the ground-state of the $n=6$ case, corresponding to the $a^{-}a^{-}c^{+}$
 tilting pattern, was neglected in our preliminary study of Ref. \cite{myself-npjCM2022}.}. Our findings demonstrate the crucial role played
 by J-T distortions in the thickness-dependent structural transitions, as well as their connection to an emergent symmetry-breaking between
 Mn sublattices within the $a^{-}a^{-}a^{-}$ tilting pattern.

%%%%%%%%%%%%%%%%%%%%%%%%%%%%%%%%%%%%%%%%%%%%%%%%%%%%%%%%%%%%%%%%
%%%%%%%%%%%%%%%%%%%%%%       Figure       %%%%%%%%%%%%%%%%%%%%%%
\begin{figure}[htb]
\centering
	\includegraphics[trim = 0 0 0 0,width=1.0\linewidth]{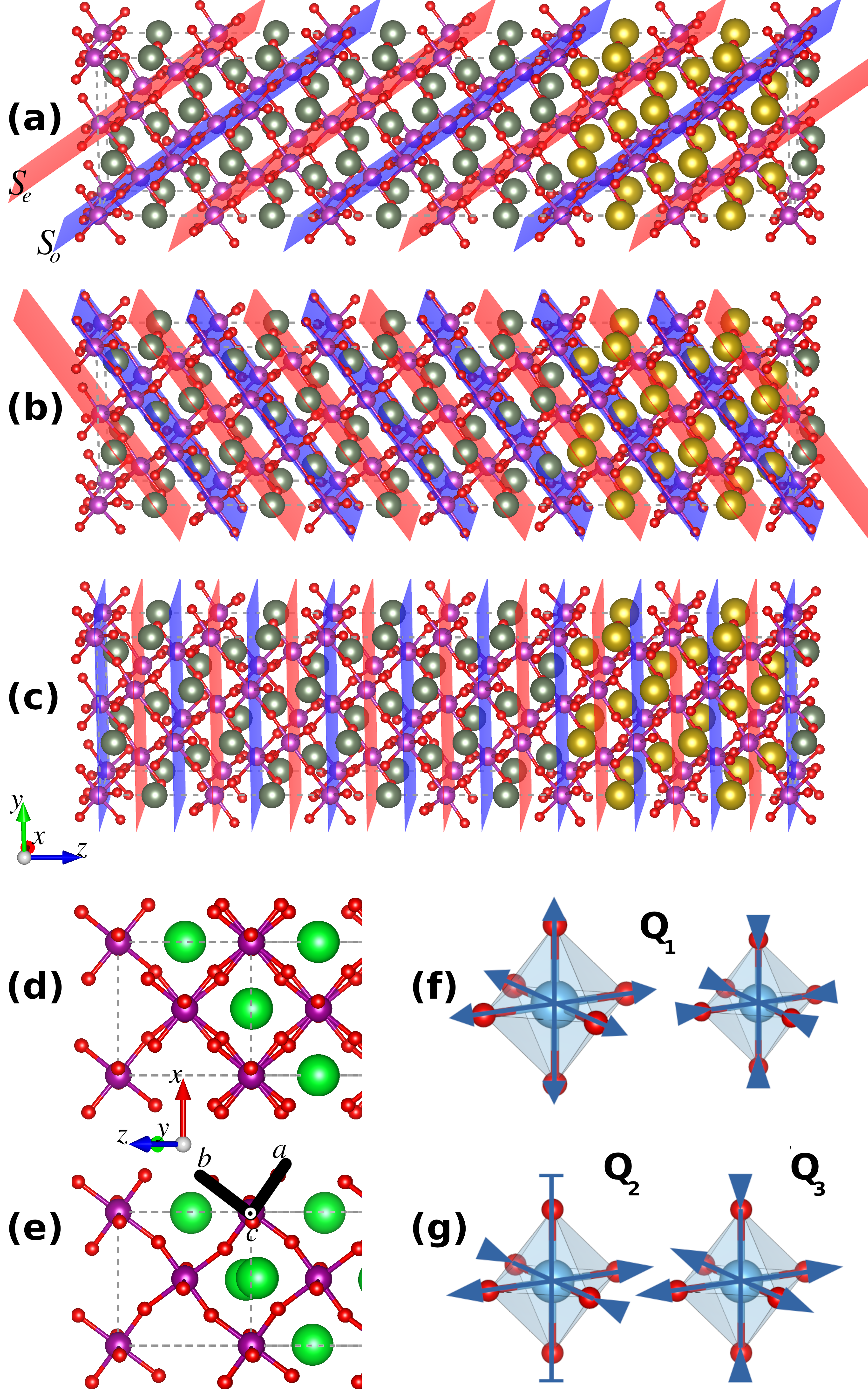}
	\caption{Sketch of the structure of the superlattice, illustrating the magnetic	orders and the tilting systems
	considered. The AFM A-type, C-type and G-type orders are illustrated in (a), (b) and (c), respectively; blue and
	red planes highlight the two spin channels. The La, Sr, Mn and O atoms are represented in dark green, yellow,
	purple and red,	respectively; in (d) and (e), A-site cations (La/Sr) are light green, whereas (f) and (g) show
	generic	transition metal and its octahedral cage. The $x$, $y$ and $z$ (the superlattice direction of growth)
	are along the crystallographic directions $(a/\sqrt{2},-b/\sqrt{2},0)$,	$(a/2,b/2,-c/\sqrt{2})$ and
	$(a/\sqrt{3},b/\sqrt{3},c/\sqrt{3})$, with axes in red, green and blue, respectively. The tilts along the $c$
	crystallographic direction can be out-of-phase (d) and in-phase	(e). The main distortions considered in this
	study are the v-b $Q_1$ (f) and the J-T $Q_2$ and $Q_3$ (g), with formulas defined in ref.\ \cite{myself-npjCM2022}.}
\label{fig:sketch}
\end{figure}
%%%%%%%%%%%%%%%%%%%%%%%%%%%%%%%%%%%%%%%%%%%%%%%%%%%%%%%%%%%%%%%%

\section{Methods and Models}
 Density functional theory (DFT) calculations are performed using the projector-augmented wave method as implemented in the Vienna {\itshape{ab-initio}}
 simulation package (VASP) \cite{kresse-PRB.54.11169,kresse-PRB.59.1758}; the generalized gradient approximation (GGA) in the Perdew–Burke–Ernzerhof
 parametrisation \cite{perdew-PRL.77.3865,perdew-PRLerratum.78.1396} is adopted. We used the pseudopotentials for Sr, La, Mn and O which treat explicitly
 10, 11, 15 and 6 electrons, respectively and we chose an energy cutoff on the plane-wave of \SI{500}{\electronvolt}. The sampling of the Brillouin Zone
 is performed with $\Gamma$-centred k-meshes of $7\times 4\times 1$, $7\times 4\times 2$ and $7\times 4\times 3$ for the $n=6$, $n=4$ and $n=2$ superlattices,
 respectively. In combination with it, a Gaussian smearing of \SI{10}{\milli\electronvolt} is used (except for the density of states (DOS), for which the
 tetrahedron method was adopted). An energy tolerance of \SI{1e-6}{\electronvolt} and of \SI{1e-7}{\electronvolt} is adopted for the electronic loop during
 the structural optimisation and the calculations of the electronic properties, respectively. Structures are considered relaxed with forces within
 \SI{5e-3}{\electronvolt\per\angstrom}. The optimised lattice constant of \SI{3.860}{\angstrom} is in agreement with a previous study \cite{myself-npjCM2022}.
 The Mn-3d states \cite{mellan-PRB.92.085151} are better described using the on-site repulsive correction via the rotationally invariant DFT+$U$ approach
 \cite{liechtenstein-PRB.52.r5467}, with Hubbard and Hund parameters $U =$ \SI{3.8}{\electronvolt} and $J =$ \SI{1.0}{\electronvolt}, respectively.
 These values are in line with previous work on (001)-oriented superlattices \cite{nanda-PRB.81.224408,nanda-PRL.101.127201,myself-npjCM2022}. Moreover,
 we have recently demonstrated that calculations performed with the parameter-free metaGGA strongly constrained and appropriately normed (SCAN) functional
 \cite{sun_jw-PRL.115.036402} yield very similar results, confirming that the specific choice of $U$ and $J$ is not crucial for our scope
 \footnote{The metaGGA SCAN functional was proven to be very successful in describing the physics of magnetic oxides \cite{varignon-ncomm2019}}. The A-type,
 C-type and G-type AFM orders, shown in FIG.\ \ref{fig:sketch}(a-c), are compared to the FM order (not illustrated). In the following, we will refer to these
 orders as A-AFM, C-AFM and G-AFM, while we will use the term spin to indicate the spin magnetic moments. Full structural relaxation was performed to obtain
 the lattice parameters, based on energy and stress tensor minimisation \cite{myself-npjCM2022}. Crystallographic directions are defined by the Mn-O bonds
 and referred to as $a$, $b$ and $c$; Cartesian axes $x$, $y$ and $z$ are the two in-plane directions and the out-of-plane direction of the superlattice,
 respectively; crystallographic and Cartesian directions are illustrated in FIG.\ \ref{fig:sketch}(d,e). In the (111) orientation, all three crystallographic
 directions $a$, $b$ and $c$ have both in-plane and out-of-plane components, as opposed to (001) orientation, where two crystallographic axes lie on the
 in-plane direction while the third axis coincides with the out-of-plane direction (the direction of growth). The rotation of the octahedra around the $c$
 axis is determined by the Mn-O-Mn angles. As octahedra are stacked along the $c$ direction, these rotations can be either out-of-phase -- FIG.\
 \ref{fig:sketch}(d) -- or in-phase -- FIG.\ \ref{fig:sketch}(e) -- accounting for the $a^{-}a^{-}a^{-}$ and $a^{-}a^{-}c^{+}$ tilting systems, respectively
 (notice the superimposed positions of O lying along the $a$ and $b$ axes in FIG.\ \ref{fig:sketch}(e) as opposed to FIG.\ \ref{fig:sketch}(d)). The
 volume-breathing (v-b) distortion $Q_1$ is illustrated in FIG.\ \ref{fig:sketch}(f); the volume-conserving J-T distortions $Q_2$ and $Q_3$ are illustrated
 in FIG.\ \ref{fig:sketch}(g). For the layer-resolved magnitudes of Q$_1$, Q$_2$ and Q$_3$, we use the same notation formalised by van Vleck and employed in
 previous work \cite{schmitt-PRB.101.214304,myself-npjCM2022,vanvleck-JChemPhys1939}; layered-resolved charge and spin distributions are computed according
 to Bader theory \cite{bader-AccChemRes1985}.
 
 Superlattices are built from the $Pnma$ and $R\bar{3}c$ bulk structures, featuring $a^{-}a^{-}c^{+}$ and $a^{-}a^{-}a^{-}$ tilting
 systems, respectively. Describing the former tilting system requires a doubling of the in-plane periodicity with respect to the latter
 one, for (111)-oriented superlattices; this corresponds to 2 formula units per layer. For an accurate comparison of energy with the
 same spacing of reciprocal lattice points, we nevertheless model the structure with $a^{-}a^{-}a^{-}$ tilting systems in the same
 $Pnma$ supercell. The La and Sr slabs alternate with thickness $2n$ and $n$, respectively, along the (111) direction, and $n$ takes
 the values $2$, $4$ and $6$. Only the $n=6$ case is shown in FIG.\ \ref{fig:sketch} as the other are perfectly analogous. Odd values
 of the thickness would result in structural and magnetic frustration, and are not treated in the current study.
 
 Finally, the images of the structures are produced with VESTA JP-Minerals \cite{VESTA-JACr2011}, and the analysis of the electronic properties is performed with
 the aid of the post-processing code VASPKIT \cite{VASPKIT-1908.08269}.

\section{Results}
 We start with the thickness-dependent structural and magnetic hierarchy illustrated in TABLE \ref{toten_strc-magn:tab}. For all
 values of $n$, the magnetic ground state is FM, whereas the most competitive AFM order is A-AFM, which highlights the driving role
 played by LaMnO$_3$; further, the C-AFM and G-AFM follow in this order. A transition between the $a^{-}a^{-}a^{-}$ tilting pattern
 and the $a^{-}a^{-}c^{+}$ tilting pattern characterises the structural order, see TABLE \ref{toten_strc-magn:tab}, and determines
 some intriguing property, as we shall see below. For a comparison with LaMnO$_3$ and SrMnO$_3$ in the bulk, we remind the reader
 to the results reported in reference \onlinecite{myself-npjCM2022} (supplemental material), where the A-AFM and the G-AFM orders
 are preferred to the FM order by \SI{8.3}{meV} per formula unit and \SI{99.2}{meV} per formula unit in bulk LaMnO$_3$ and bulk
 SrMnO$_3$, respectively.

%%%%%%%%%%%%%%%%%%%%%%%%%%%%%%%%%%%%%%%%%%%%%%%%%%%%%%%%%%%%%%%%%%%
%%%%%%%%%%%%%%%%%%%%%%%%       Table       %%%%%%%%%%%%%%%%%%%%%%%%
\begin{table}[ht]
	\centering
	\caption{Relative energy of various magnetic states and tilting patterns, labelled with respect to the space group, of (LaMnO$_3$)$_{2n}\vert$(SrMnO$_3$)$_{n}$ superlattices
	with $n=6,4,2$. Values are given in \SI{}{meV} per formula unit and with respect to the ground state (GS) for any given $n$. The dashes for $n=2$ in the $Pnma$ structure
	($a^{-}a^{-}c^{+}$ tilting system) indicate that our calculations never converged to this arrangement, but always transitioned to the $R\bar{3}c$ structure ($a^{-}a^{-}a^{-}$
	tilting system).}
	\label{toten_strc-magn:tab}
	\begin{tabular}{r|cccc|cccc}
%      FM &   (0.21)   &     GS     \\
%  A-type &  (50.73)   &   47.11    \\
%  C-type &  (80.50)   &   80.98    \\
%  G-type &  163.33    &  109.41    \\
		  &      \multicolumn{4}{c|}{$Pnma$}                 &        \multicolumn{4}{c}{$R\bar{3}c$}            \\
                  &   FM     &   A-AFM    &   C-AFM    &   G-AFM     &     FM     &   A-AFM    &   C-AFM    &   G-AFM    \\
		 \hline
           $n=6$  &   GS     &   26.46    &   43.79    &   72.16     &    7.62    &   37.53    &   53.84    &   95.60    \\
		 \hline
           $n=4$  &   GS     &   38.49    &   67.21    &   95.95     &    1.25    &   38.60    &   59.92    &  137.46    \\
		 \hline
           $n=2$  &  ---     &    ---     &    ---     &    ---      &     GS     &   47.11    &   80.98    &  109.41    \\
               
 \end{tabular}
\end{table}
%%%%%%%%%%%%%%%%%%%%%%%%%%%%%%%%%%%%%%%%%%%%%%%%%%%%%%%%%%%%%%%%%%%

\subsection{Structural properties and magnetic hierarchy}
\begin{figure}[ht]
\centering
      \includegraphics[trim = 20 0 0 0,width=0.95\linewidth]{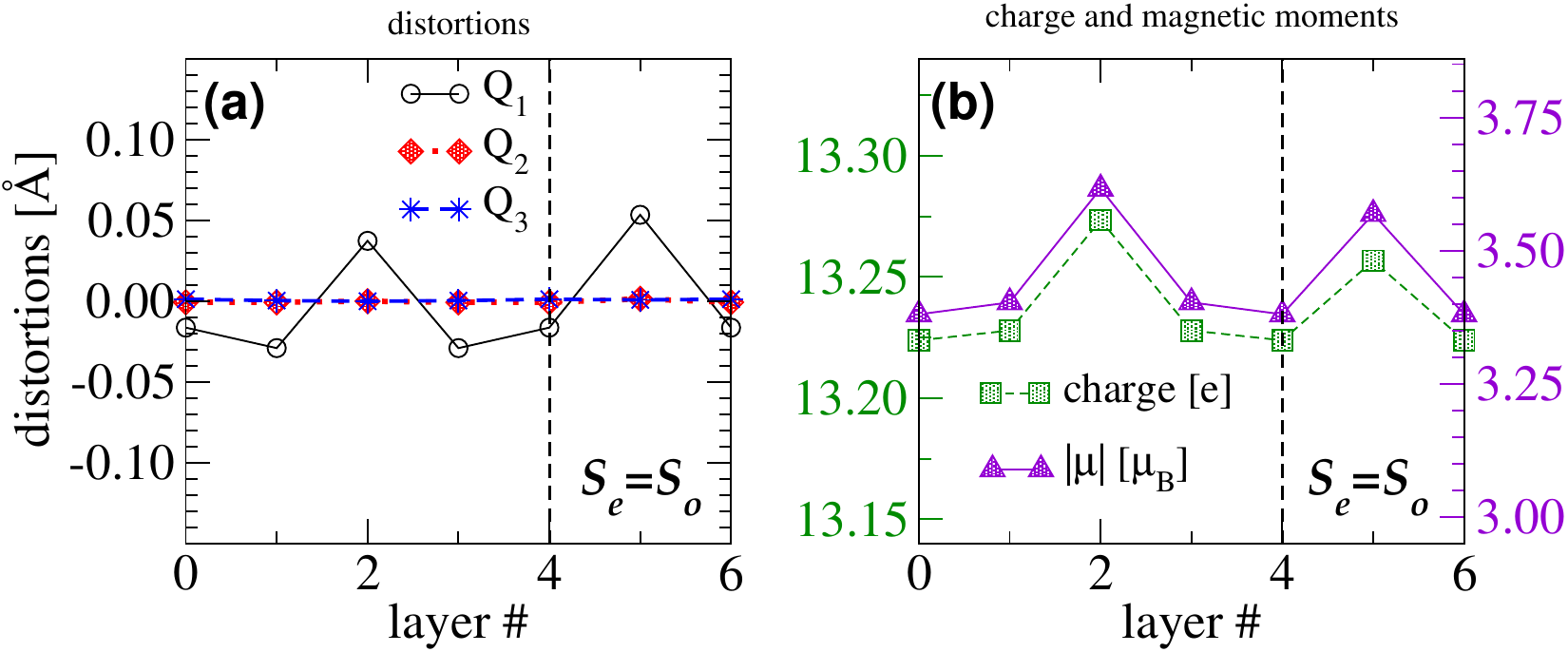}
        \caption{Layer-resolved van Vleck distortions (a) and charge and
        spin distributions (b) of the $n=2$ superlattice, FM solution. The
	$S_o$ and $S_e$ sublattices show similar properties.}
\label{fig:42f3-vvd-chsp}
\end{figure}

 Naturally, the results for $n=2$ are the closest to bulk La$_{2/3}$Sr$_{1/3}$MnO$_3$, in line with recent measurements of the magnetic and transport properties
 of (111)-oriented (LaMnO$_3$)$_{2}\vert$(SrMnO$_3$)$_{1}$ superlattices \cite{wang_zz-PSSb2022}. For such a small thickness, the $a^{-}a^{-}a^{-}$ tilting pattern
 is the ground state, whereas it is not even possible to stabilize the $a^{-}a^{-}c^{+}$ tilting pattern as a metastable state. The octahedral distortions are virtually
 null, mirrored by a homogeneous distribution of charge and spin, see FIG.\ \ref{fig:42f3-vvd-chsp}. Despite the difference in the chemical environment around the
 Mn between the SrMnO$_3$ region and the LaMnO$_3$ region, the Mn charge remains the same. In this scenario of valence states, the octahedra in the SrMnO$_3$ region
 tend to be larger than those in the LaMnO$_3$ region. Because of the symmetry, the two Mn atom lying on the same layer are equivalent. With a notation which is
 explained in detail below, we indicate this fact by '$S_e = S_o$'. The magnetic order is FM, with a rather large energy gain with respect to the competing AFM
 orders, see TABLE \ref{toten_strc-magn:tab}. We further notice that in ref.\ \onlinecite{wang_zz-PSSb2022} the `SrMnO$_3$ layer' has a Mn surrounded by Sr on one
 side and La on the other, whereas in the $n=2$ case of the current study there is a Mn layer surrounded by Sr on both sides.

%%%%%%%%%%%%%%%%%%%%%%%%%%%%%%%%%%%%%%%%%%%%%%%%%%%%%%%%%%%%%%%%
%%%%%%%%%%%%%%%%%%%%%%       Figure       %%%%%%%%%%%%%%%%%%%%%%
%%%%%%%%%%%%%%%%%%%%%%%%%%%%%%%%%%%%%%%%%%%%%%%%%%%%%%%%%%%%%%%%

 As we increase the thickness of the superlattice, the ground-state remains FM for all thicknesses and tilting patterns, see TABLE \ref{toten_strc-magn:tab}. For a thickness larger than
 those we considered, we expect to recover bulk properties for both regions, namely a FM to A-AFM transition in the LaMnO$_3$ region and a FM to G-AFM transition in the SrMnO$_3$ region.
 A simulation of the mixed A-AFM/G-AFM order was performed to verify this hypothesis, finding an energy of \SI{26.4}{\milli\electronvolt} per formula unit above the ground state, meaning
 that such mixed order is still unfavourable at $n=6$. Additionally, a mixed A-AFM/FM order was simulated, finding an energy of \SI{26.8}{\milli\electronvolt} per formula unit above the
 ground state. Therefore, with the A-AFM order in the LaMnO$_3$ regions and FM, G-AFM and A-AFM in the SrMnO$_3$ region (compare also with TABLE \ref{toten_strc-magn:tab}), a negligible
 energy cost is found for a FM-AFM transition in the SrMnO$_3$ region. This is due to the close and subtle competition between FM and AFM exchange coupling revealed in our previous work
 \cite{myself-npjCM2022}. This competition also suggests that the FM to G-AFM transition in the SrMnO$_3$ region is likely to be preceded by local spin-flips, which is consistent with the
 fact that bulk SrMnO$_3$ is a wide-gap band insulator.
 
%%%%%%%%%%%%%%%%%%%%%%%%%%%%%%%%%%%%%%%%%%%%%%%%%%%%%%%%%%%%%%%%
%%%%%%%%%%%%%%%%%%%%%%       Figure       %%%%%%%%%%%%%%%%%%%%%%
\begin{figure}[hb]
\centering
        \includegraphics[trim = 20 0 0 0,width=0.95\linewidth]{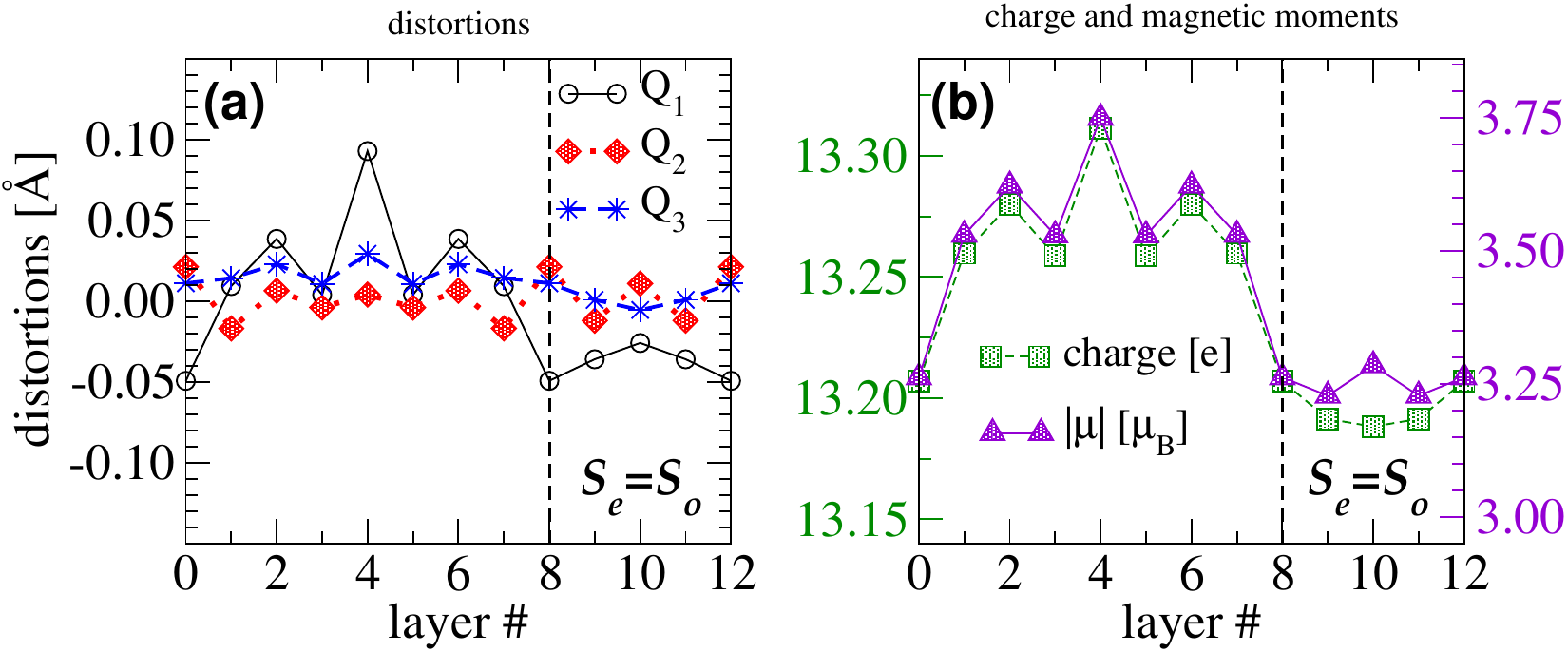}
        \caption{Layer-resolved van Vleck distortions (a) and charge and
        spin distributions (b) of the $n=4$ superlattice, FM solution. The
	$S_o$ and $S_e$ sublattices show similar properties.}
\label{fig:84f3-vvd-chsp}
\end{figure}
%%%%%%%%%%%%%%%%%%%%%%%%%%%%%%%%%%%%%%%%%%%%%%%%%%%%%%%%%%%%%%%%
 
%%%%%%%%%%%%%%%%%%%%%%%%%%%%%%%%%%%%%%%%%%%%%%%%%%%%%%%%%%%%%%%%
%%%%%%%%%%%%%%%%%%%%%%       Figure       %%%%%%%%%%%%%%%%%%%%%%
\begin{figure}[hb]
\centering
        \includegraphics[trim = 20 0 0 0,width=0.95\linewidth]{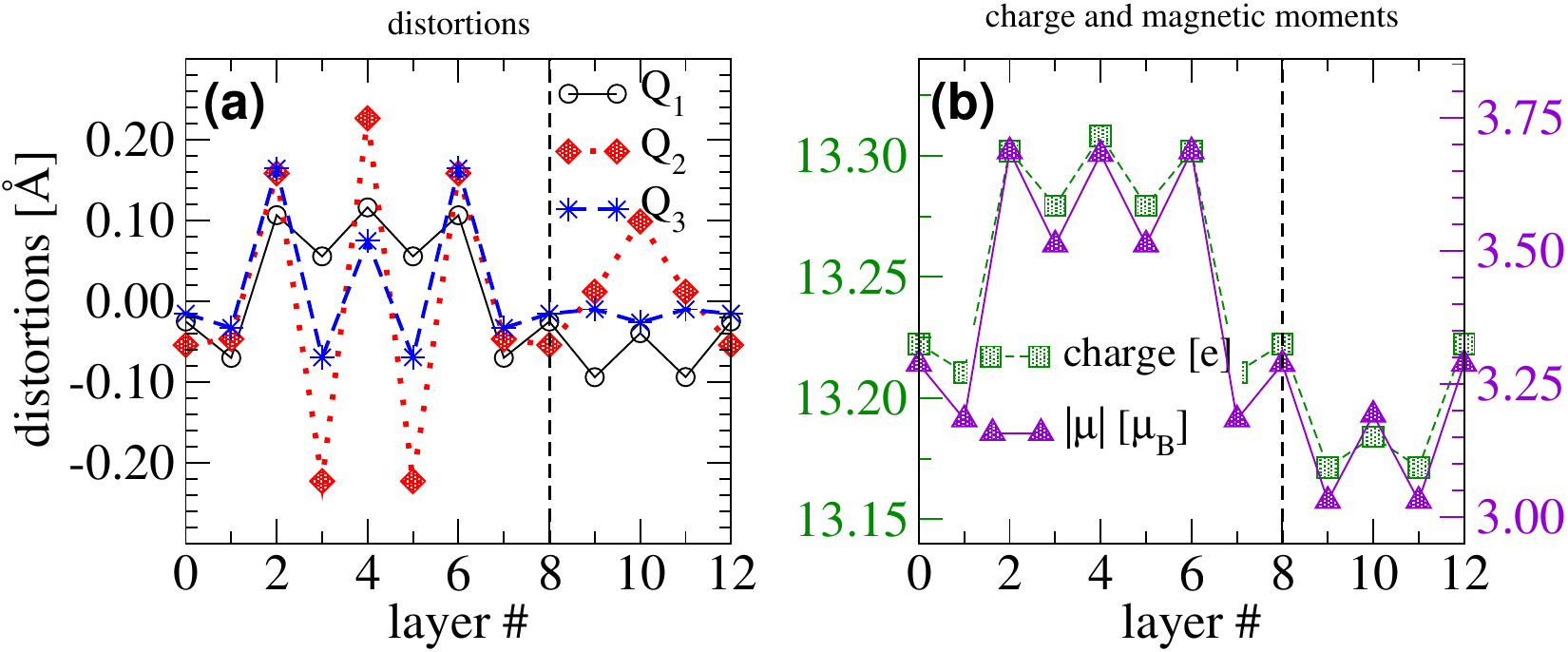}
        \caption{Layer-resolved van Vleck distortions (a) and charge and spin
	distributions (b) of the $n=4$ superlattice, A-AFM solution. The $S_o$
	and $S_e$ sublattices show similar properties.}
\label{fig:84f2-vvd-chsp}
\end{figure}
%%%%%%%%%%%%%%%%%%%%%%%%%%%%%%%%%%%%%%%%%%%%%%%%%%%%%%%%%%%%%%%%

% we observe a variety of structural effects, which will be analysed below, but
 Moving to the structural analysis, and as mentioned above, TABLE \ref{toten_strc-magn:tab} shows that increasing thickness from $n=2$ to $n=4$ induces a
 change of tilting pattern, from $a^{-}a^{-}a^{-}$ to $a^{-}a^{-}c^{+}$. The thicker superlattice allows for more variability in the plots of the lattice
 distortions and the charge/spin distributions across the layers, illustrated in FIG.\ \ref{fig:84f3-vvd-chsp}. With hindsight, we can group the Mn sites
 into two distinct sublattices, depicted as alternating blue and red (001) planes in FIG.\ \ref{fig:sketch}(a) and labeled as `$S_o$' and `$S_e$', respectively
 \footnote{These sublattices reflect the different character of the magnetic coupling in the A-AFM ground-state of bulk LaMnO$_3$, since the inter-atomic
 exchange coupling is FM within each plane but AFM between planes.}. Despite these sublattices are different by symmetry the results obtained for $n=4$ show
 a quasi-degeneracy, for all magnetic orders. Therefore, only one set of curves is reported in FIG.\ \ref{fig:84f3-vvd-chsp}, for a clearer visualization.
 These data show that, for $n=4$, the $Q_1$ v-b distortion in the FM ground state dominates over the J-T distortions, still quenched. The suppression of the
 J-T distortions here is not a mere consequence of the tilting system \cite{varignon-PRB.100.035119,varignon-PRR.1.033131} and the small LaMnO$_3$ thickness,
 but is an effect of the magnetic degrees of freedom. In fact, the J-T distortions emerge in the A-AFM solution, as illustrated by FIG.\ \ref{fig:84f2-vvd-chsp}.
 This is particularly evident for $Q_2$, as it is linked to the orbital order along the [001] planes, likely promoting FM coupling therein. These findings are
 consistent with the fact that J-T distortions are necessary for the formation of the A-AFM order
 \cite{pickett_we-PRB.53.1146,geck-NJP2004,pavarini-PRL.104.086402,schmitt-PRB.101.214304}.
 Going back to the analysis of FIG.\ \ref{fig:84f3-vvd-chsp}, volume, charge and spin oscillate in the LaMnO$_3$ region, featuring a peak in the innermost layer;
 the v-b $Q_1$ distortion is mirrored by the layered-resolved charge/spin distribution, as previously reported for the $a^{-}a^{-}a^{-}$ tilting pattern ($n=6$)
 \cite{myself-npjCM2022}. In the centre of the SrMnO$_3$ region, volume and spin reach a peak, whereas the charge varies more smoothly. This demonstrates the
 complex interplay between the various degrees of freedom which prevents a simple picture based on the assumption of a homogeneous charge transfer.
 
%%%%%%%%%%%%%%%%%%%%%%%%%%%%%%%%%%%%%%%%%%%%%%%%%%%%%%%%%%%%%%%%
%%%%%%%%%%%%%%%%%%%%%%       Figure       %%%%%%%%%%%%%%%%%%%%%%
\begin{figure}[ht]
\centering
	\includegraphics[trim = 20 0 0 0,width=0.95\linewidth]{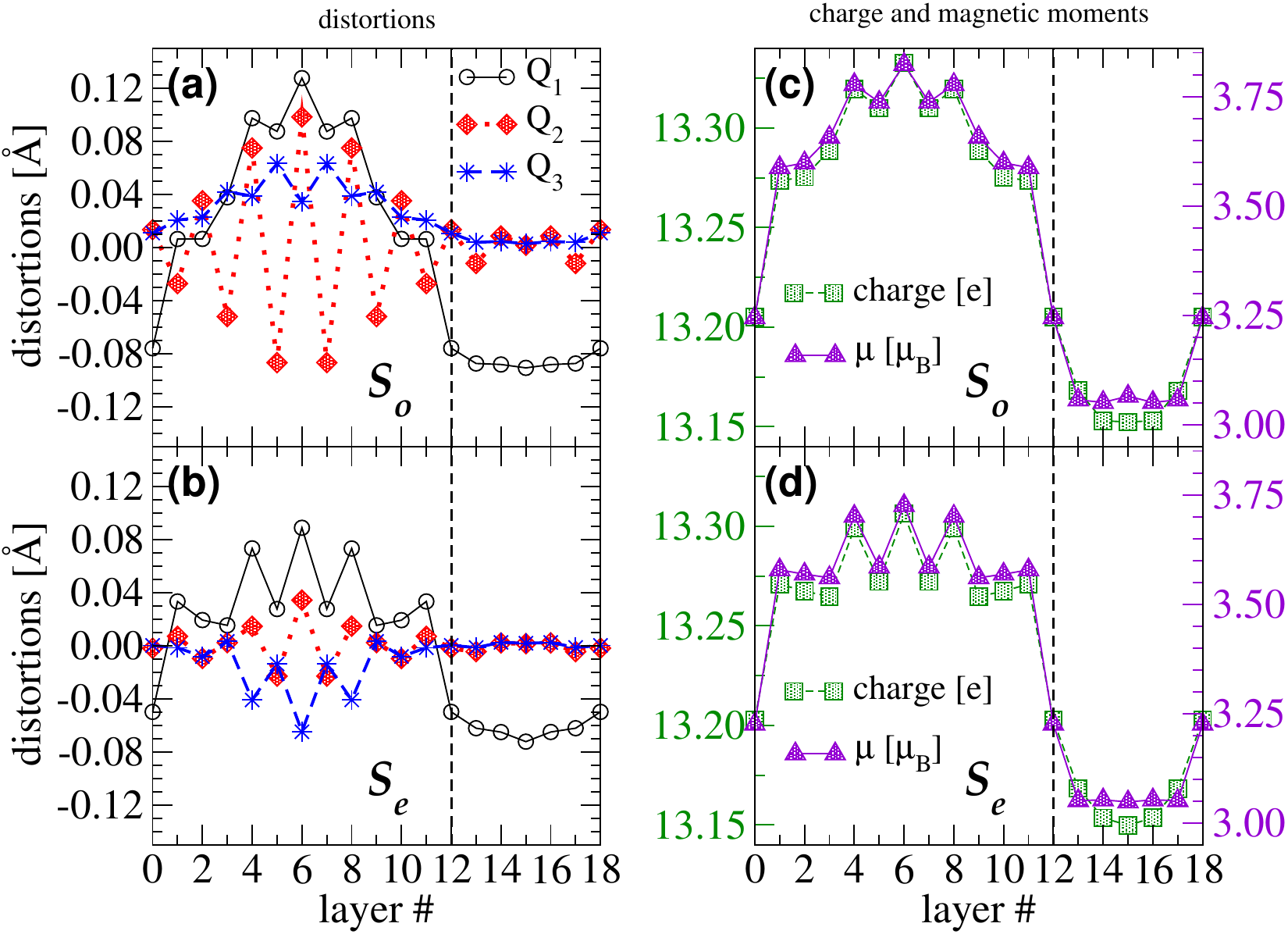}
	\caption{Layer-resolved structural and electronic/magnetic properties
	of the ground state of the $n=6$ superlattice, the FM solution with the
	$a^{-}a^{-}c^{+}$ tilting pattern: van Vleck distortions in sublattice
	$S_o$ (a) and $S_e$ (b); charge and spin distributions in sublattices
	$S_o$ (c) and $S_o$ (d).}
\label{fig:f3-vvd-chsp}
\end{figure}
%%%%%%%%%%%%%%%%%%%%%%%%%%%%%%%%%%%%%%%%%%%%%%%%%%%%%%%%%%%%%%%%

%%%%%%%%%%%%%%%%%%%%%%%%%%%%%%%%%%%%%%%%%%%%%%%%%%%%%%%%%%%%%%%%
%%%%%%%%%%%%%%%%%%%%%%       Figure       %%%%%%%%%%%%%%%%%%%%%%

\begin{figure}[hb]
\centering
	\includegraphics[trim = 20 0 0 0,width=0.95\linewidth]{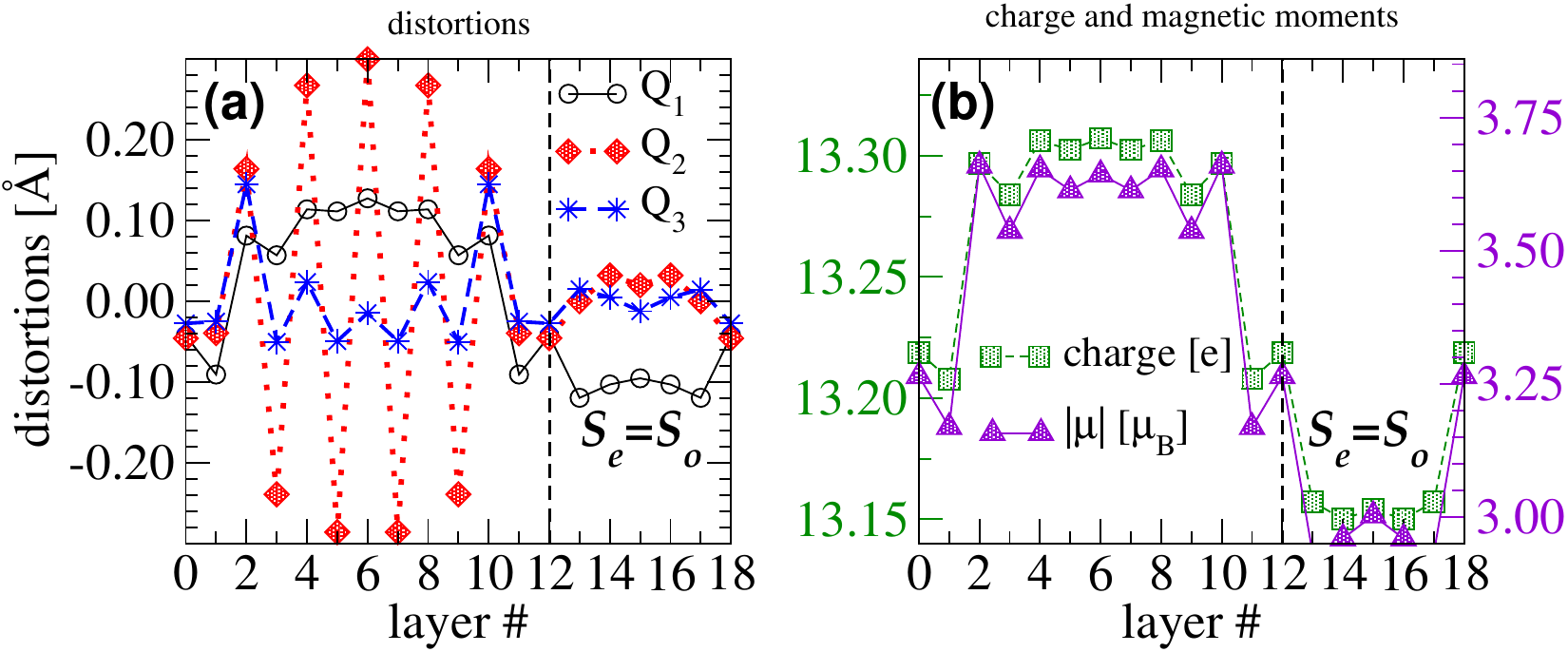}
	\caption{Layer-resolved van Vleck distortions (a) and charge and
	spin distributions (b) of the $n=6$ superlattice, A-AFM solution.
	The $S_o$ and $S_e$ sublattices show similar properties.}
\label{fig:f2-vvd-chsp}
\end{figure}
%%%%%%%%%%%%%%%%%%%%%%%%%%%%%%%%%%%%%%%%%%%%%%%%%%%%%%%%%%%%%%%%

%%%%%%%%%%%%%%%%%%%%%%%%%%%%%%%%%%%%%%%%%%%%%%%%%%%%%%%%%%%%%%%%
%%%%%%%%%%%%%%%%%%%%%%       Figure       %%%%%%%%%%%%%%%%%%%%%%
\begin{figure}[ht]
\centering
	\includegraphics[trim = 20 0 0 0,width=0.95\linewidth]{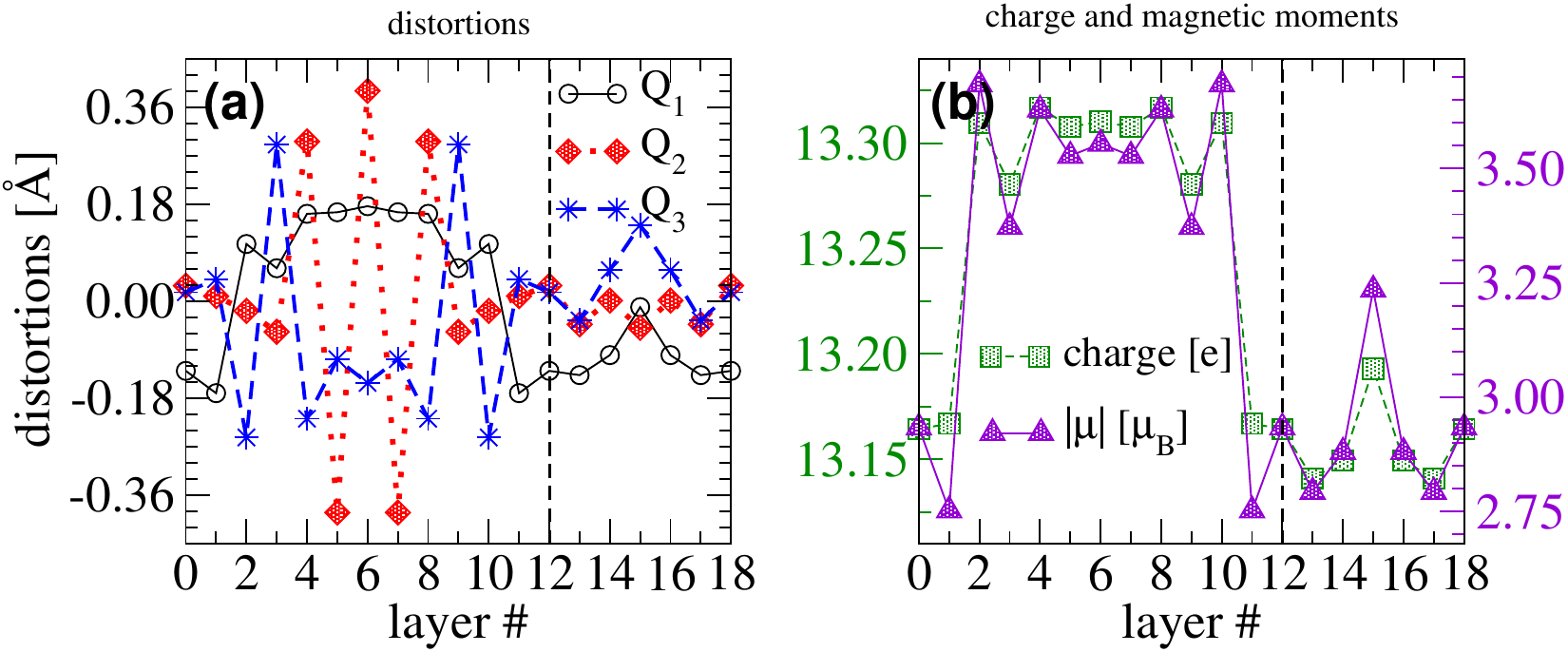}
	\caption{Layer-resolved van Vleck distortions (a) and charge and
	spin distributions (b) of the $n=6$ superlattice, C-AFM solution.
	The $S_o$ and $S_e$ sublattices show similar properties.}
\label{fig:f1-vvd-chsp}
\end{figure}
%%%%%%%%%%%%%%%%%%%%%%%%%%%%%%%%%%%%%%%%%%%%%%%%%%%%%%%%%%%%%%%%

 We now move to the data for $n=6$, which is the largest thickness we study. Figure \ref{fig:f3-vvd-chsp} shows the distribution of layered-resolved van Vleck distortions, charge and
 spin of the FM ground state. In contrast with the $n=4$ case, a dramatic difference arises between the sublattices $S_o$ and $S_e$. The v-b $Q_1$ distortion is still mirrored by the
 layered-resolved charge/spin distribution -- compare FIG.\ \ref{fig:f3-vvd-chsp}(a) with FIG.\ \ref{fig:f3-vvd-chsp}(c) and FIG.\ \ref{fig:f3-vvd-chsp}(b) with FIG.\
 \ref{fig:f3-vvd-chsp}(d). Moreover, we now have marked J-T distortions for the FM ground state. The J-T $Q_2$ distortion is obviously accompanied by orbital order
 \cite{rao_cnr-JPCB2000,geck-NJP2004}.
 The largest contributions to $Q_2$ arise mainly from the $S_o$ sublattice, see FIG.\ \ref{fig:f3-vvd-chsp}(a), while these modes seem quenched in the $S_e$ sublattice, see FIG.\
 \ref{fig:f3-vvd-chsp}(b).

 The relative weight of the $Q_1$ mode is particularly large in the $S_e$ sublattice, while it is comparable to $Q_2$ for $S_o$. Moreover, $Q_1$ exhibits
 oscillations in the $S_e$ sublattice, while it varies smoothly in the $S_o$ sublattice. Therefore, also for $Q_1$ we observe a qualitative difference
 across the two sublattices, which is connected to the charge distribution. The larger charge in the $S_o$ sublattice, see FIG.\ \ref{fig:f3-vvd-chsp}(c),
 points to a larger La-Sr valence separation therein and suggests a propensity of the $S_o$ sublattice to restore the bulk-like orbital order and the J-T
 distortions, by withstanding the interfacial charge transfer from LaMnO$_3$ to SrMnO$_3$. Such valence separation is crucial for the emergence of mixed
 structural features, because a valence closer to $3+$ drives the $e_{g}$ occupation closer to 1/2, prompting J-T distortions. On the other hand, opposite
 values of the $Q_3$ distortion and different charge states in the LaMnO$_3$ region for the two sublattices promote a FM coupling along (001) within the
 Goodenough-Kanamori model \cite{goodenough-PhysRev1955,kanamori-JPandCS1959,goodenough-InterPub1963}. Note that the hopping between two sublattices occurs
 not along the same layer (same $z$ value), but between adjacent layers.

% The $Q_3$ oscillations are out-of-phase in the two sublattices, suggesting that octahedra in the two different (001)-oriented planes are tall or short, promoting hopping between
% $d_{3z^2-r^2}$ orbitals with different occupancy. This pattern is mirrored by the orbital projections shown in FIG.\ \ref{fig:f3-orb}: the eigenstates in the centre of the LaMnO$_3$
% region have more (less) occupancy in the $S_o$ ($S_e$) sublattice, and also the orbital mixing follows the same trend. With respect to the Goodenough-Kanamori rules
% \cite{goodenough-PhysRev1955,kanamori-JPandCS1959,goodenough-InterPub1963}, such scenario favours a FM coupling between the $S_o$ and $S_e$ sublattices.

 % A-type, n=6 vV distortions and chg/mgm
 We proceed to the analysis of the A-AFM order for $n=6$. The van Vleck distortions and the charge/spin distribution are illustrated in FIG.\ \ref{fig:f2-vvd-chsp}. The curves for $S_o$
 and $S_e$ sublattices are virtually equivalent, and therefore only one set of curves is shown. Analogously to the $n=4$ case, the $Q_2$ distortion in the A-AFM solution is larger than
 that in the FM ground state and arise in the LaMnO$_3$ region, unsurprisingly. The $Q_3$ distortion oscillates between positive and negative values, with a small amplitude, and peaks at
 the interfaces. In contrast with the FM solution, $Q_1$ changes abruptly at the interface but does not show oscillations within either bulk region, see FIG.\ \ref{fig:f2-vvd-chsp}(a);
 again, the $Q_1$ distortion is mirrored by charge/spin oscillations, see FIG.\ \ref{fig:f2-vvd-chsp}(b).

%%%%%%%%%%%%%%%%%%%%%%%%%%%%%%%%%%%%%%%%%%%%%%%%%%%%%%%%%%%%%%%%
%%%%%%%%%%%%%%%%%%%%%%       Figure       %%%%%%%%%%%%%%%%%%%%%%

\begin{figure*}[ht]
\centering
	\includegraphics[trim = 0 0 0 0,width=0.25\linewidth]{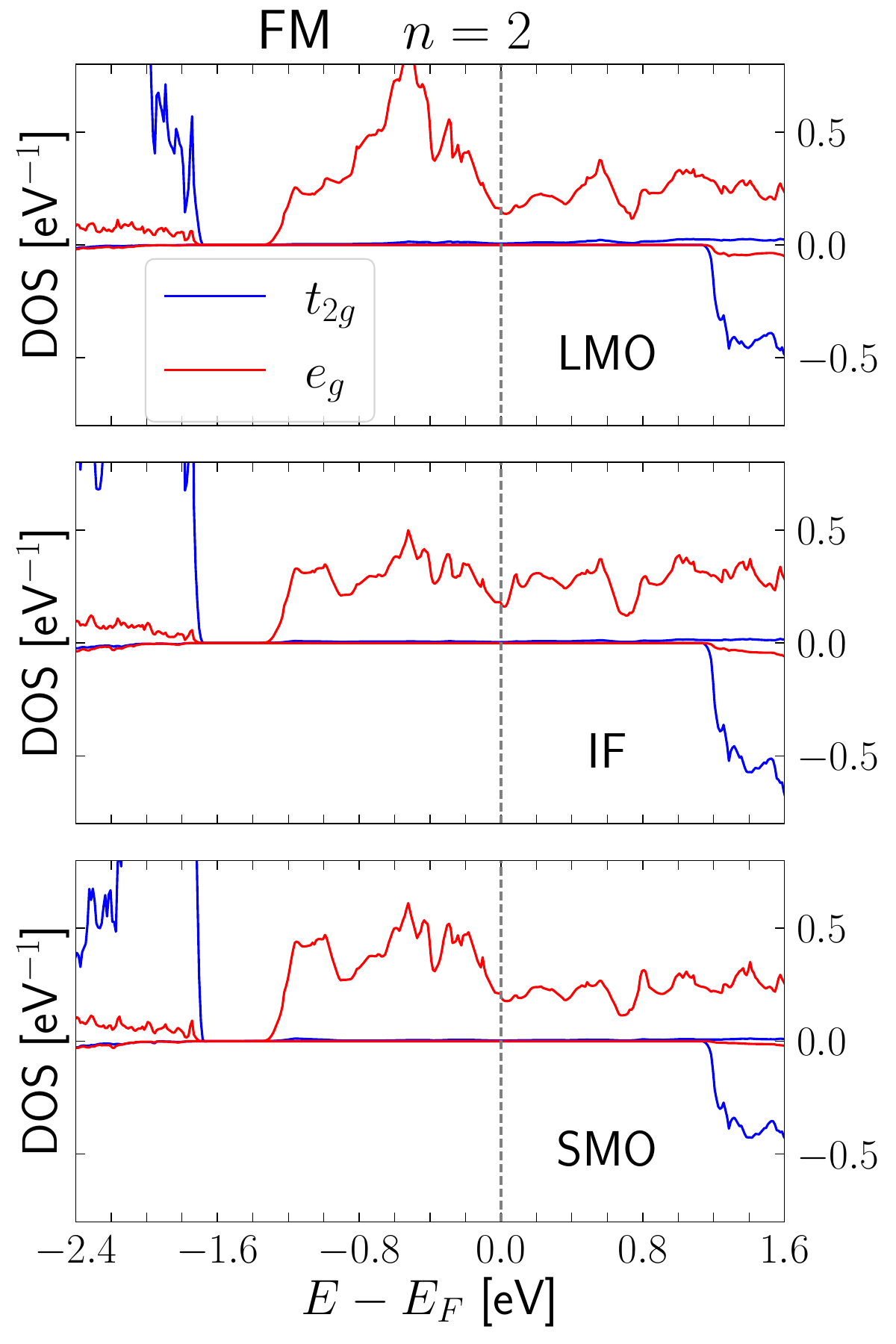}
	\includegraphics[trim = 0 0 0 0,width=0.25\linewidth]{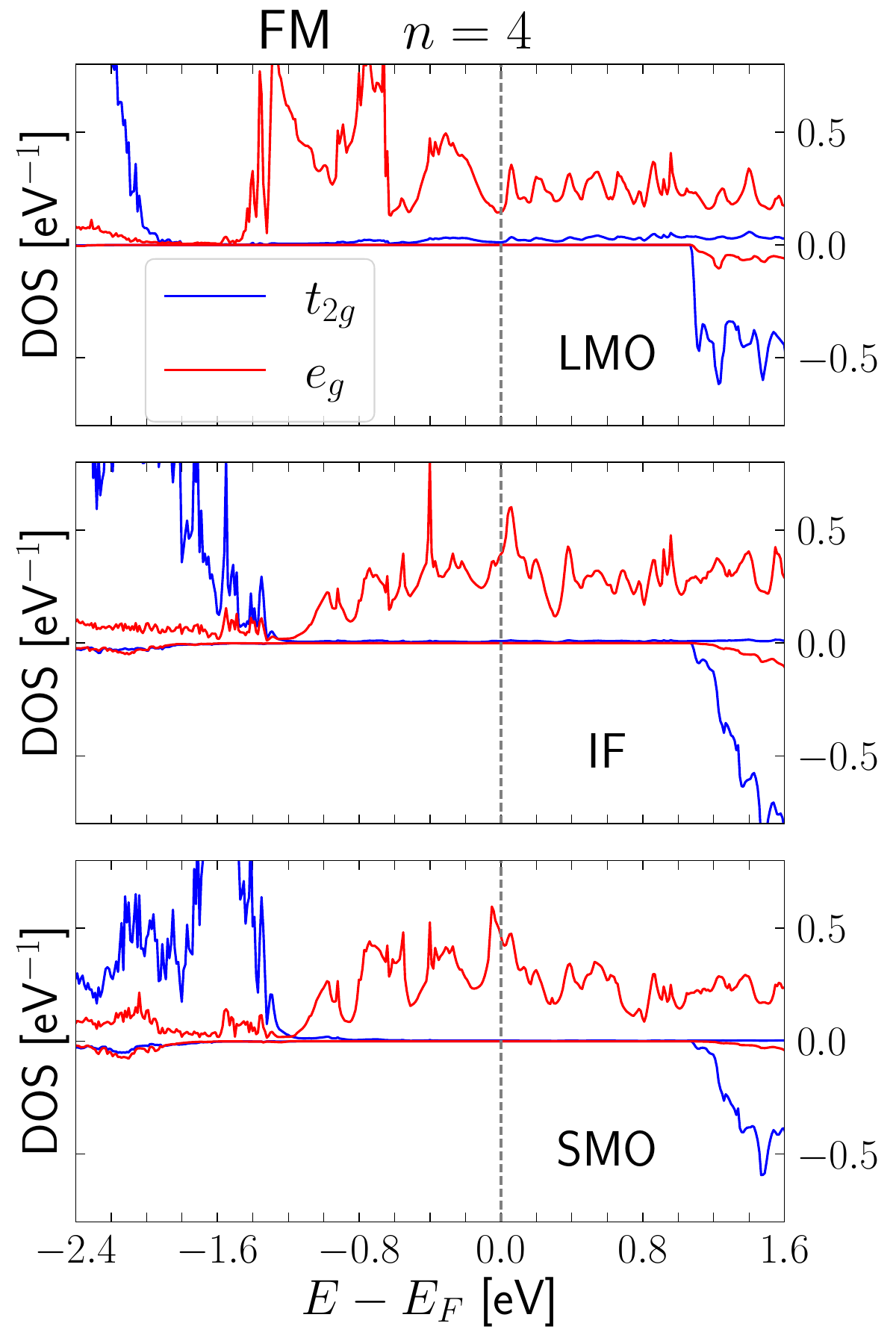}
	\includegraphics[trim = 0 0 0 0,width=0.25\linewidth]{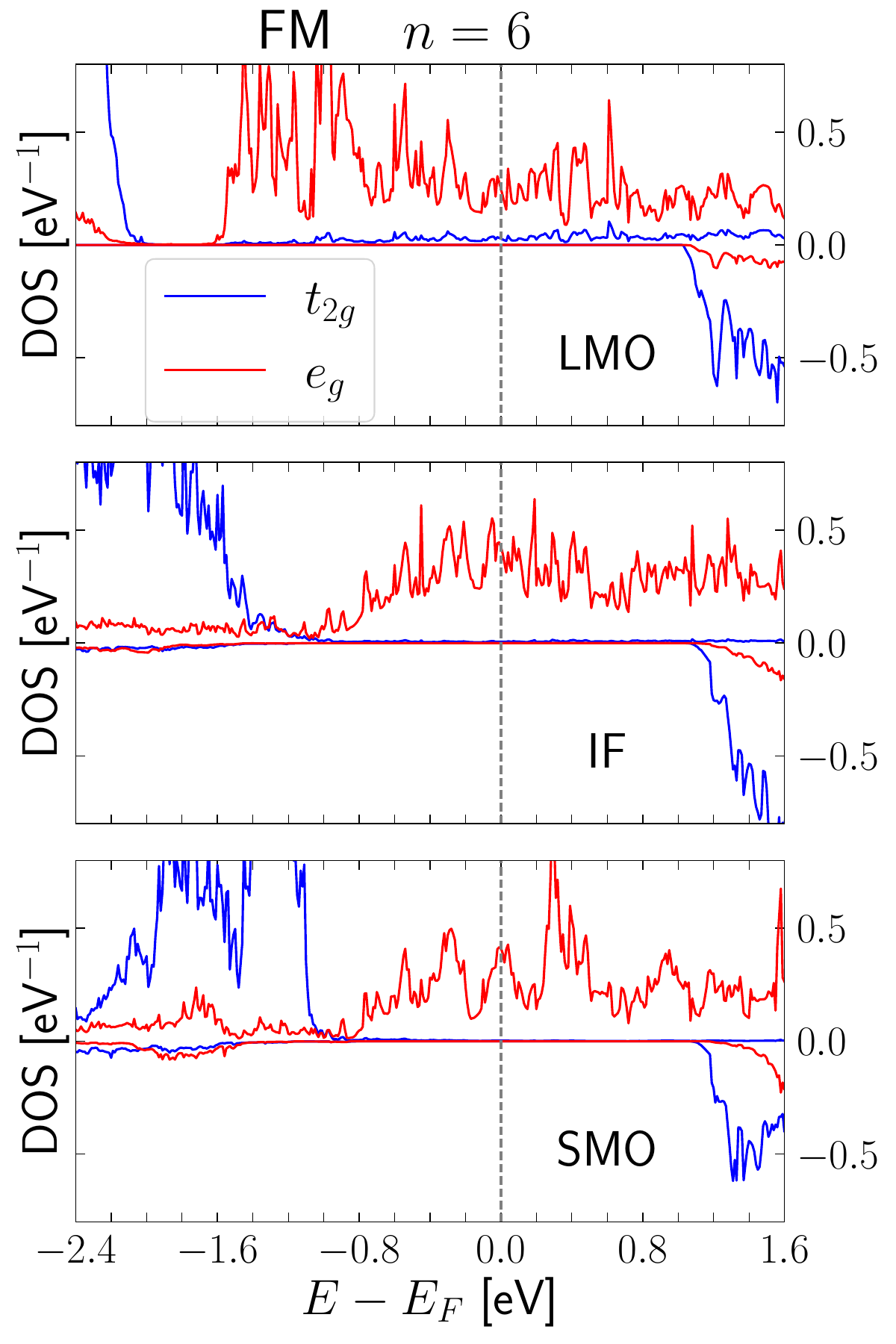}\\
\vspace{0.25cm}
	\includegraphics[trim = 0 0 0 0,width=0.25\linewidth]{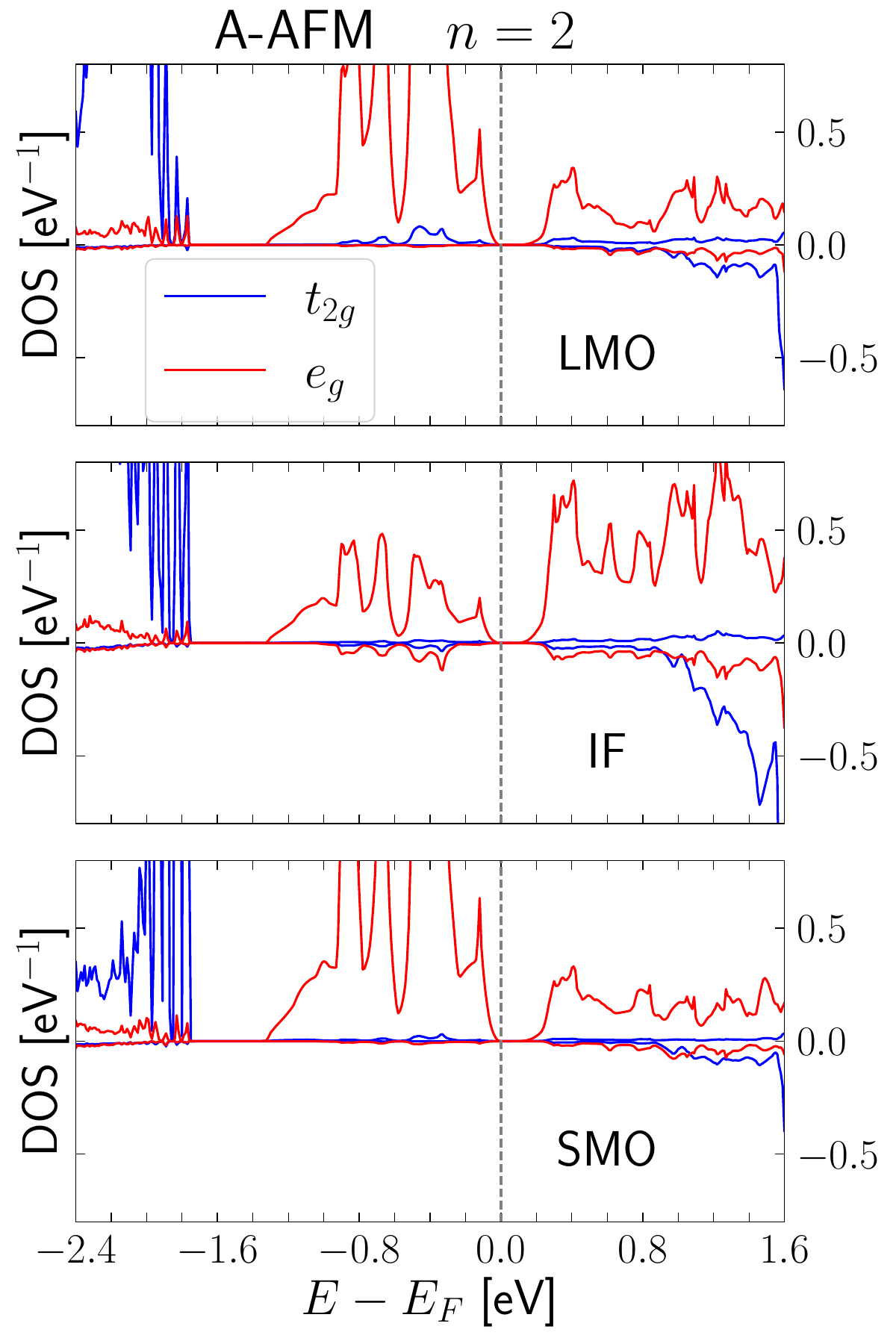}
	\includegraphics[trim = 0 0 0 0,width=0.25\linewidth]{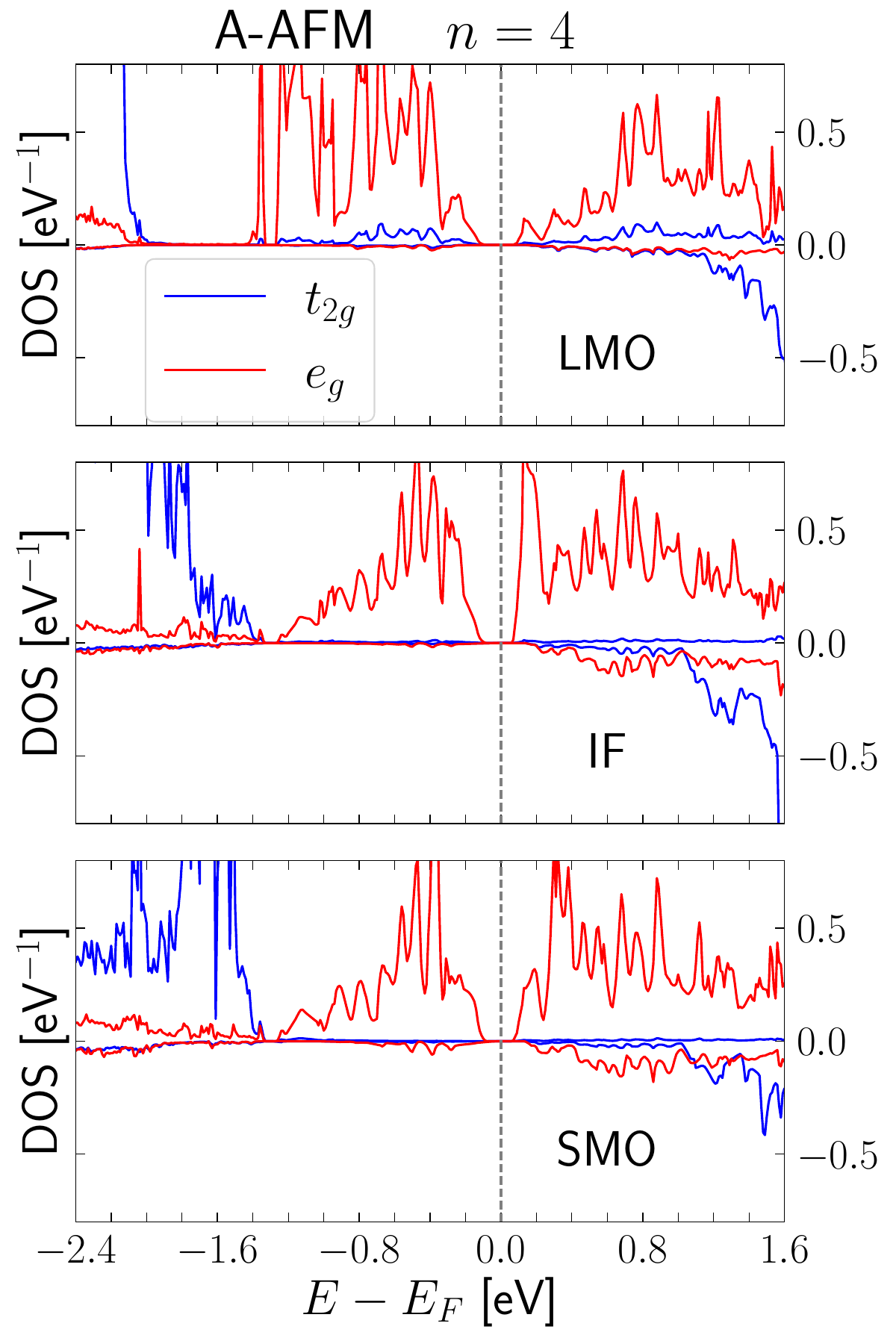}
	\includegraphics[trim = 0 0 0 0,width=0.25\linewidth]{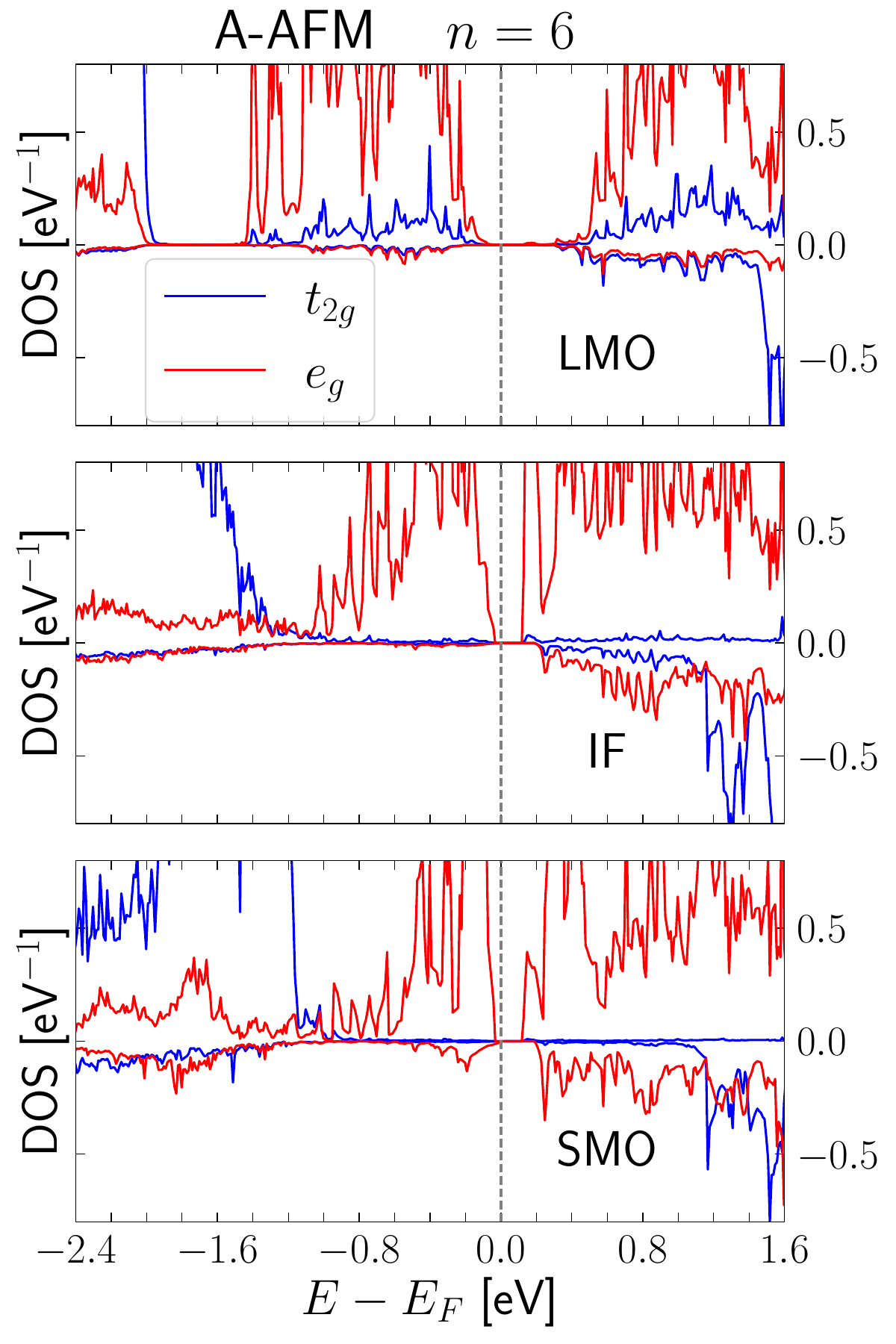}
	\caption{Layer-resolved PDOS of the $n=2$, $n=4$ and $n=6$ superlattices,
	FM and A-AFM solution. For all systems, we represent the central layer
	of the LaMnO$_3$ region (LMO), the layer at the interface (IF) and the
	central layer of the SrMnO$_3$ region (SMO).}
\label{fig:dos}
\end{figure*}
%%%%%%%%%%%%%%%%%%%%%%%%%%%%%%%%%%%%%%%%%%%%%%%%%%%%%%%%%%%%%%%%
 
 Further calculations, for example the analysis of the structural distortions of the C-AFM, reveal that the quasi-equivalence between $S_e$ and $S_o$ sublattices
 does not depend on the type of AFM order; therefore, as illustrated in FIG.\ \ref{fig:f1-vvd-chsp}, we show only one set. While in the FM solution single-spin
 sublattices display a broken structural symmetry, if the spin is compelled to adopt two distinct states the lattice degrees of freedom adjust and relax to a
 single-phase configuration.

\subsection{Electronic properties}
 % DOS results withholding that J-T distortions alone will induce a metal-insulator transition and/or a FM-AFM transition.
 For every thickness, the FM solution features a half-metallic state which persists across all the layers of the superlattice, as shown by the projected density-of-states (PDOS) in FIG.\
 \ref{fig:dos}. The A-AFM solution is instead fully insulating with a band gap across the $e_{g}$ states. Such gap is constant throughout the superlattice for $n=2$ and $n=4$, whereas it
 is enhanced in the LaMnO$_3$ region for $n=6$. In particular, it amounts to \SI{\sim 0.16}{\electronvolt}, \SI{\sim 0.22}{\electronvolt} and \SI{0.56}{\electronvolt} for $n=2$, $n=4$ and
 $n=6$ (LaMnO$_3$ region), respectively. These values are well below the value of \SI{\sim 1.2}{\electronvolt} calculated for bulk LaMnO$_3$, confirming that the relaxation to an insulating
 AFM phase (bulk-like) may happen only at a larger thickness. This reflects a spatially-extended charge transfer between Sr and La regions, which is also evident from the charge densities
 (data not shown) and the plots of the Bader charges. Overall, these features are fully consistent with recent data for (111)-oriented LaAlO$_3\vert$SrTiO$_3$ superlattices
 \cite{song_k-nnano2018}. In connection with the aforementioned FM-AFM transition, expected at large $n$, this charge distribution will cease extending and eventually recede when the system
 becomes insulating.

 Furthermore, we observe that the character of the bands in both the FM and the A-AFM solution slightly changes with thickness. In fact, a gap between the
 $t_{2g}$ states and $e_{g}$ states exists in all regions (LaMnO$_3$, interface, SrMnO$_3$) for $n=2$, whereas it is present only in the LaMnO$_3$ region
 for $n=4$ and $n=6$, see the DOS in FIG.\ \ref{fig:dos}; such trend is seen for both FM and A-AFM solutions. On the other hand, a residual $t_{2g}$-$e_{g}$
 mixing is observed in the LaMnO$_3$ region, which can reflect the octahedral distortions as seen in FIGs.\ \ref{fig:42f3-vvd-chsp}-\ref{fig:f2-vvd-chsp}.

 We notice that our computational approach may neglect effects that are detrimental to the half-metallic character we predict in this family of superlattices. For example, we do not
 include spin-orbit coupling nor we investigate the occurrence of non-collinear magnetis; this latter would mix the two spin channels and decrease spin polarisation of the carriers
 \cite{fetzer_r-SciRep2015,dimarco-PRB.97.035105}. Moreover, we do not include explicit many body effects, which may lead to non-quasiparticle states forming inside the minority-spin
 band gap \cite{katsnelson-RMP.80.315}. While these effects go beyond the scope of the present study, which we expect to be unaffected in terms of structural and magnetic hierarchy,
 as well as excitation spectra, one should also stress that deviations from a full spin polarisation may become larger and much more relevant in transport properties, depending on
 the type of measurement \cite{nadgorny_b-PRB.63.184433}.

\section{Discussion and Conclusions}
 The main results of the current study are the evolution of properties with thickness -- showing that Jahn-Teller distortions kick in before the metal-insulator transition or the
 FM-AFM transition -- and the relation of the structural and spin degrees of freedom -- showing a sublattice separation of structural phases in the spin-degenerate FM ground state
 and that the sublattices become (quasi) degenerate in structural properties when the spin degeneracy is lifted. The former result highlights the primary role played by the J-T
 distortions originating from the LaMnO$_3$ region of the superlattice; the latter points to a symmetry-dependent separation -- it occurs only in presence of the $a^{-}a^{-}c^{+}$
 tilting pattern -- which is realised on either of the lattice or spin degrees of freedom. The fact that J-T distortions in the FM phase are sizeable for $n=6$, but dramatically
 quenched for $n=4$, is interpreted as a precursor of the transition to a different tilting pattern, which happens for $n=2$.

 An interesting question arising from our study is on the emergence of the separation of J-T and v-b sublattices and its connection to magnetism. Our most
 plausible explanation is drawn in the light of an antagonism between the A-AFM magnetic order of the LaMnO$_3$ -- promoted by J-T and the $a^{-}a^{-}c^{+}$
 tilting pattern -- and the FM order of the superlattice -- promoted by strain and charge transfer. While in the $a^{-}a^{-}a^{-}$ tilting pattern the J-T
 distortions are naturally suppressed by symmetry \cite{varignon-PRB.100.035119,malyi_o-APR2020}, as the thickness increases and the tilting pattern of bulk
 LaMnO$_3$ is adopted, the J-T distortions appear in the AFM solutions. For the largest thickness considered, while the FM order is still preferred, the tilting
 pattern tends to promote J-T distortions but cannot maximise them via an insulating A-AFM order because such state is not energetically competitive. Thus, the
 system adopts a mixed configuration: in one sublattice, the J-T are stronger and in the other they are weaker and overshadowed by the v-b. This interpretation
 also suggests why the two sublattices remain degenerate for $n=4$ even in the FM solution: there is not enough room for the above-mentionied competition to
 develop along the direction of growth.

 A broader connection to the analysis above is provided by existing measurements and models for bulk LaMnO$_3$. On one hand, Raman spectra \cite{baldini_m-PRL.106.066402} and magneto-transport
 measurements \cite{baldini_m-PNAS2015} show the occurrence of a phase separation, where J-T regions are sided by regions with no J-T, under a moderate hydrostatic pressure. On the other hand,
 \textit{ab-initio} calculations show a (hidden) competition between van Vleck modes $Q_2$ (J-T) and $Q_1$ (v-b) \cite{schmitt-PRB.101.214304}. We advance the hypothesis that in (111)-oriented
 mixed-valent superlattices, a combination of strain and charge transfer parallel the moderate hydrostatic pressure causing the aforementioned phase separation. Further investigations based on
 scanning transmission electron microscopy (STEM), could verify the actual realisation and character of structural features -- such as octahedral tilts -- which are deeply linked to electronic
 and magnetic properties according to our predictions. In addition, orbital occupations (at the interfaces) may be investigated using X-ray magnetic linear dichroism (XMLD) in reflectivity. We
 also note that epitaxial strain may tune the structural phase separation observed in the current study, as it does not occur with the $a^{-}a^{-}a^{-}$ tilting pattern which is favoured by a
 range of substrates.

% ok, but why in a-a-c+ and not in a-a-a-? vol-breathing & Hund's vs J-T and Mott's
 Finally, our results support the importance of symmetry in the description of electronic properties and electronic correlations, as recent research work
 \cite{varignon-ncomm2019,varignon-PRB.100.035119,varignon-PRR.1.033131,malyi_o-APR2020} has highlighted. Since v-b (J-T) distortions are linked to the
 emergence of Hund (Mott) correlations \cite{mazin-PRL.98.176406,varignon-PRB.100.035119}, further research on these and similar superlattices is expected
 to unveil the strong intertwining between Mott and Hund physics. In this sense, similar systems, often showing Hund-driven charge disproportionation, are
 nickelates \cite{mazin-PRL.98.176406,li_d-Nature2019,chang_j-EPJb2020,lechermann-PRX.10.041002,wang_y-PRB.102.161118,kang_bk-npjQM2023} and ruthenates
 \cite{georges-AnnuRevCMP2013,tyler-PRB.58.R10107,wang_sc-PRL.92.137002}.

 Concluding, we presented an \textit{ab-initio} study on (111)-oriented (LaMnO$_3$)$_{2n}\vert$(SrMnO$_3$)$_{n}$ superlattices with $n=2,4,6$. All studied systems exhibit a robust
 half-metallic FM order, persistent across all the layers. We observe a crossover between bulk La$_{2/3}$Sr$_{1/3}$MnO$_3$ and bulk LaMnO$_3$ with varying thickness, where the J-T
 distortions play a crucial role. In the $Pnma$ structure, the FM GS consists of two sublattices with qualitatively different van Vleck distortions and charge/spin distributions.
 These sublattices become quasi-degenerate for all AFM orders, which are also accompanied by growing J-T distortions.
 %This highlights the importance of the LaMnO$_3$ lattice degrees of freedom, since SrMnO$_3$ is a band insulator and only electronic and magnetic degrees of freedom are active.
 %These degrees of freedom, and its resulting features, need a minimum number of units to be activated.
 These findings highlight the complex and fascinating relationship between lattice, charge, orbital and spin degrees of freedom in (111)-oriented
 manganite superlattices, and underscore their potential for novel functionalities and applications.

%\section{Acknowledgments}
\begin{acknowledgments}
	We are thankful to I.\ I.\ Mazin, A.\ Edstr\"om and A.\ Akbari for valuable discussions. The computational resources were provided by the Korean
	Institute of Science and Technology Information (KISTI) national supercomputing centre (Project No. KSC-2022-CRE-0358),) by the National Academic
	Infrastructure for Supercomputing in Sweden (NAISS) and the Swedish National Infrastructure for Computing (SNIC) at the Center for High Performance
	Computing (PDC) in Stockholm, Sweden, partially funded by the Swedish Research Council through grant agreements no. 2022-06725 and no. 2018-05973;
	we additionally appreciate the computational support from the University of York High-Performance Computing service, Viking, and the Research Computing
	team. F.\ C., H.-S.\ K. and I.\ D.\ M. acknowledge financial support from the National Research Foundation (NRF) funded by the Ministry of Science
	of Korea (Grants Nos. 2022R1I1A1A01071974, 2020R1C1C1005900 and 2020R1A2C101217411, respectively). H.-S.\ K.\ acknowledges additional support from
	the international cooperation program managed by the National Research Foundation of Korea (NRF-2023K2A9A2A12000317). I.\ D.\ M.\ acknowledgess
	financial support from the European Research Council (ERC), Synergy Grant FASTCORR, Project No. 854843. This research is also part of the project
	No. 2022/45/P/ST3/04247 co-funded by the National Science Centre and the European Union’s Horizon 2020 research and innovation programme under the
	Marie Sk\l odowska-Curie grant agreement no.\ 945339. For the purpose of Open Access, the author has applied a CC-BY public copyright licence to any
	Author Accepted Manuscript (AAM) version arising from this submission.
\end{acknowledgments}


%%%%%%%%%%%%%%%%%%%%%%%%%%%%%%%%%%%%%%%%%%%%%%%%%%%%%%%%%%%%%%%%%%%%%%%%%%%
\end{document}